\documentclass[twocolumn,showpacs,preprintnumbers,amsmath,amssymb,prb]{revtex4}
\usepackage{graphicx}
\usepackage{dcolumn}
\usepackage{bm}
\usepackage{wasysym}
\usepackage{textcomp}
\usepackage{epstopdf}
\usepackage{pxfonts}
\usepackage{txfonts}
\usepackage{tipa}
\usepackage{fontenc}
\usepackage{color,soul}
\begin{document}

\preprint{APS/123-QED}

\title{Anamalous positive magnetization signal at the onset of normal to superconductor transition in single crystal Nb sphere}
\author{P. D. Kulkarni$^{*1,2}$, S. S. Banerjee$^{3}$, H. Suderow$^{2}$, H. Takeya$^{4}$, S. Ramakrishnan$^{1}$}
\affiliation{$^{1}$Department of Condensed Matter Physics and Materials
Science, Tata Institute of Fundamental Research, Homi Bhabha Road,
Colaba, Mumbai 400005, India.}
\affiliation{$^{2}$Laboratorio de Bajas Temperaturas, Departamento de Fisica de la Materia Condensada, Instituto de Cinecia de Materiales Nicolas Cabrera, Facultad de Ciencias University Autonoma de Madrid, E-28049, Madrid, Spain}
\affiliation{$^{3}$Department of Physics, Indian Institute of Technology Kanpur, Kanpur 208016, India}
\affiliation{$^{4}$National Institute of Materials Science, Ibaraki 305-0047, Japan}

\date{\today}
\begin{abstract}
We report the observation of positive magnetization signal at the onset of superconducting transition in isothemral and isomagnetic measurements in single crystal Niobium sphere. The PME displays two regimes, viz., low applied magnetic fields ($H \leq 4$\,Oe) and higher fields ($H > 120$\,Oe), and the PME signal is absent at the intermediate magnetic fields. Subsequently, the isothermal magnetization data shows the presence of positive signal at the superconductor to normal transition. Using this data we are able to plot the ($H_{c3}$-$T_{c3}$) line, at which the positive magnetization signal nucleates, in the phase diagram of single crystal Nb.

\end{abstract}

\pacs{74.25.Ld, 74.25.Ha, 74.25.Op}


\maketitle

The positive magnetization signal below the normal to superconductor transition was observed in several samples of high-$T_c$ oxides, as well as conventional low $T_c$ elements and alloys \cite{Khomskii,Braunisch,Svedlindh,Reidling,Sigrist,Li,Thompson,Pust,Geim,Das}. In the former systems, the response is designated as anti-Meissner\cite{Svedlindh} or Wohlleben effect\cite{Wohlleben}, however, considering several similar observations in conventional superconductors, the response can be termed as parmagnetic Meisnner effect (PME). 

First reports of PME were in granualar Bi$_2$Sr$_2$CaCu$_2$O$_8$ \cite{Braunisch, Svedlindh}. It was shown that the field cooled (FC) magnetization signal at very low magnetic fields remain positive well below the superconducting transition temperature, instead of the usual diamagnetic response. The positive magnetization signal was linked to the existence of the spontaneous orbital currents\cite{Sigrist, Li}. A more detailed explanation invoking the granular nature of the samples was proposed based on the Josephson junctions between grains. In these models it was suggested that the cooper pairs acquire a phase $\pi$ in the tunneling process across the junctions. If several grains form a current loop with odd number of such junctions a spontaneous current develops giving rise to the orbital moment. A network of randomly distributed odd number of $\pi$ junctions may behave as an orbital glass with paramagnetic response. This was also intricately linked to the $d$ wave superconducting order parameter in high $T_c$ materials. In contrast, the observeration of PME in $s$-wave superconductors, with low pinning characteristics, e.g., Nb and Al disks, necessitated flux trapping and compression alongwith the role of surface superconductivity as plausible mechanisms for the observation of PME\cite{Koshelev, Khalil, Moshchalkov, Zharkov}. These processes require flux free sample edges at the onset of the superconducting transition. This would set the currents loops counterclock wise at the outer and inner surface superconducting regions. Subsequently these regions grow towards the center of the sample, compressing the flux in the interior. Such a possibility can lead to a paramagnetic signal in the magnetization measurements. 

The quantum state of the trapped flux was studied using the Ginzburg-Landau equations \cite{Moshchalkov, Zharkov} and it was shown that the instead of the Meissner state, the giant vortex state with orbital quantum number $L$ \textgreater 0 would have the lower energy at the superconduction transition. As the sample is cooled the giant vortex would successively transformed into different lower energy states, and the magnetization signal would fluctuate before settling in the diamagnetic state. Recently, in the cubic stannide, Ca$_3$Rh$_4$Sn$_{13}$ ($T_c \sim 8.35$\,K), the detailed dc and ac magnetization studies of PME were reported\cite{Kulkarni}. For this compound, the magnetization response at very low fields was fluctuacting and it could be related the presence of giant vortex state and its subsequent transformation to the Abrikosov vortex lattice predicted in the literature \cite{Li, Moshchalkov, Zharkov}. However, in these measurements the sign of the PME signal remained intriguing. If the external magnetic field is notionally negative, the diamagnetic superconducting state in the magnetization measurements gives positive signal. However, the PME signal does not change sign, and retains the same sign as the diamagnetic signal in the negative fields. In this case, the PME signal appears as superposing to the diamagnetic signal. 

The positive field cooled magnetization (PMFC) in high purity single crystal Nb sphere was reported earlier by Das et al. \cite{Das}. The results suggested that the surface superconductivity co-exist with the PMFC. However, there were no features in these experiments which could be ascribed to the metastable nature of giant vortex states in the temperature interval of the PMFC regime. The results in Ca$_3$Rh$_4$Sn$_{13}$ motivated further very low field studies, and the detailed investigation of the normal to superconductor transition in Nb sphere. Here we report some interesting features related to the PME effect in the high purity single crystal Nb sphere. We also estimated the $H_{c3}$ values at several temperatures to plot a reliable $H_{c3}$ line on the magnetic phase diagram. 

\begin{figure}[ht]
\includegraphics[width=0.45\textwidth]{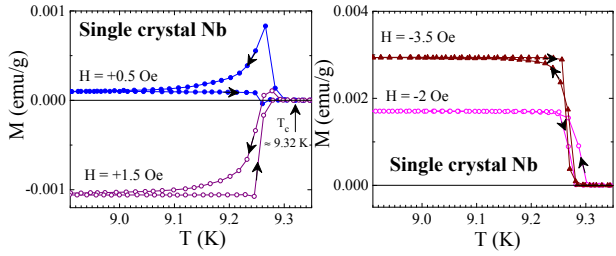}
\caption{\label{fig1}(colour online) Field cooling cool-down (FCC) and field cooling warm-up curves at different magnetic fields in single crystal Nb sphere. In panel (a), positive magnetic field curves, at 0.5 Oe and 1.5 Oe, are shown with $T_c$ marked. In panel (b), negative magnetic field curves, measured at -2 Oe and -3.5 Oe, are shown.}
\end{figure}

A single crystal sample of Nb grown under containerless condition\cite{Sung} was used. The crystal was mounted to apply magnetic field along the crystallographic [100] direction. The dc magnetization measurements were performed using a commercial SQUID-Vibrating Sample Magnetometer (Quantum Design (QD) Inc., USA, model S-VSM). In S-VSM, the sample executes a small vibration around a mean position, where the magnetic field is uniform and maximum. This avoids the possibility of the sample moving in an inhomogeneous field during the dc magnetization measurements. The remnant field of the superconducting magnet of S-VSM was carefully estimated at different stages of the experiment, using a standard paramagnetic Palladium specimen.

\begin{figure}[ht]
\includegraphics[width=0.45\textwidth]{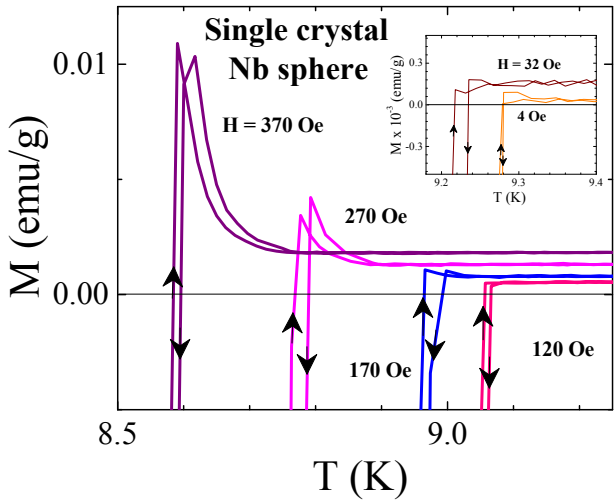}
\caption{\label{fig1}(colour online) In the main panel, we show field cooling cool-down (FCC) and field cooling warm-up (FCW) curves at intermediate magnetic fields in single crystal Nb sphere. In the inset panel, the FCC and FCW curves are shown at $H$ = 4 Oe and 32 Oe.}
\end{figure}

\begin{figure}[ht]
\includegraphics[width=0.45\textwidth]{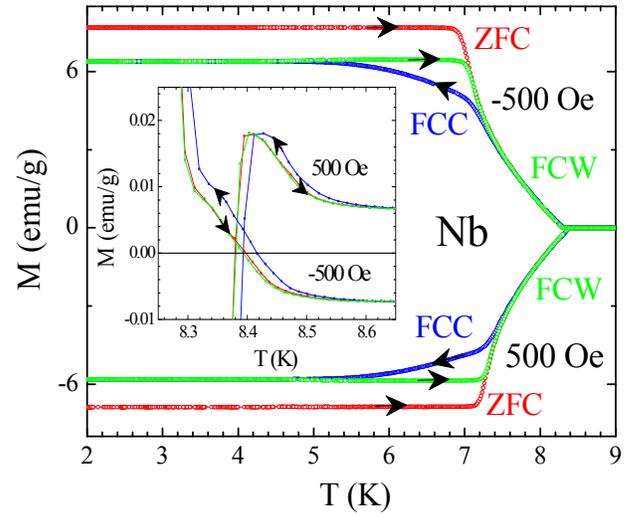}
\caption{\label{fig2}(colour online) The main panel shows the field cooled cool-down and field cooled warm-up curves at 500 Oe magnetic field on both positive and negative sides. The main panel also shows a zero field cooled warm-up measurements in 500 Oe. Inset panel shows the expanded portions of the 500 Oe curves close to the transition region at this field.}
\end{figure}

Fig.~1 shows the field cooled cool-down (FCC) and subsequent warm-up (FCW) magnetization curves are shown at small (a) positive and (b) negative magnetic fields. In panel (a) at $H$ = 0.5 Oe magnetic field, a transtion at 9.32 K is marked. At the onset of this transition, the magnetization signal is seen to rise immediately showing a sharp peak at around 9.265 K, and later the signal stabilizes to a small positive values down to the lowest temperature of 2 K in these measurements. The subsequent field cooled warm-up measurement shows the reversal of the magnetization path to split above 9 K, and producing a small negative peak at around 9.26 K before restoring the normal state magnetization signal. As the magnetic field is increased to 1.5 Oe, the positive magnetization peak suppresses compared to that in 0.5 Oe curve and the negative peak shows further shift to lower temperatures. 

Figure 2 shows the magnetization vs temperature in single crystal Nb sphere at selected magnetic fields. In the inset panel the isofield magnetization responses at $H$ = 4 Oe and 32 Oe are shown. At 4 Oe, a small increase in the magnetization can be seen before the diamagnetic response sets in below $T_c$, however at magnetic fields upto 32 Oe, the PME signal remains suppressed. In the main panel, the magnetization at $H$ = 120 Oe shows a sharp step at the onset of the superconducting transition at 9.05 K. At 170 Oe, a small peak is observed before the diamagnetic signal rapidly grows below $T_c$. Further increase in the field to 270 and 370 Oe shows the increasing peak related to the positive magnetization below the superconducting transition. 

\begin{figure}[ht]
\includegraphics[width=0.45\textwidth]{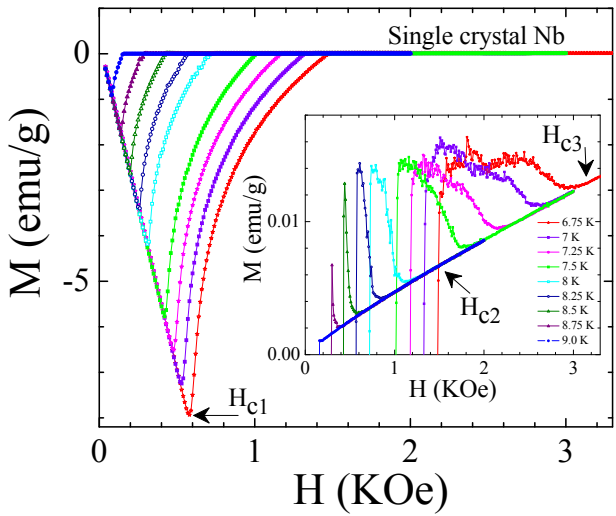}
\caption{\label{fig3}(colour online) Magnetization vs applied magnetic field at 7 K in single crystal Nb sphere. Inset in top right corner shows an expanded portion of the M-H loop highlighting the transtion region on both sides of the magnetic fields. Inset in bottom left shows highly expanded view of the transition region at the negative side. The verticle arrows in this inset show the near-periodic oscillatory positions of the magnetization signal in the transition region.}
\end{figure}

Figure 3 shows $M$ vs $T$ curves in +500 oe and -500 Oe. The FCC, FCW and ZFC curves are shown at each field. The inset shows the magnified portion of the M vs T curves in the main panel. The black arrows are marked to indicate the direction of the temperature sweep. In the inset, at the onset of the superconducting transition at 8.6 K, the magnetization is seen positive for both signs of the applied magnetic field. At $H$ = +500 Oe FCC, the magnetization signal displays a peak at 8.4 K, and later settles in the diamagnetic state below 7.4 K as seen in the main panel. The FCW curve merges with FCC curve till 4.6 K, and later both the curves again merge at 7.4 K. In the ZFC curve, the magnetization behavior is similar to FCW except slightly more diamagnetic signal. At -500 Oe FCC and FCW, the diamagnetic signals reverse their behavior and are on the positive side. However, as can be seen in the inset, the magnetization immediately below $T_c$, from 8.6 K to 8.3 K, which was associated with the paramagnetic signal at +500 Oe, now aligns with the diamagnetic signal. At the kink around 8.3 K, the diamagnetic signal in -500 Oe, takes over, and usual FCC, FCW and ZFC curves are reproduced. 

In figure 4, we show the $M$ vs $H$ loop recorded at 7 K. The forward and reverse branches are shown in blue and red curves respectively in the magnetic field range from -3000 Oe to +3000 Oe. Inset at top right corner shows the magnified portions to highlight the magnetization signal close to the transition between superconductor to normal state. In both the cases, the positive signal is observed. These are manifestations of the similar behavior in the $M$ vs $T$ curves in figure 3. At the bottom left corner we show the fine features of the broad hump in the field range from -700 Oe to -1000 Oe. The arrows are marked at the positions of the magnetization jumps. Clearly the normal to superconductor transition is marked by the oscilations in the magnetization signal.

\begin{figure}[ht]
\includegraphics[width=0.45\textwidth]{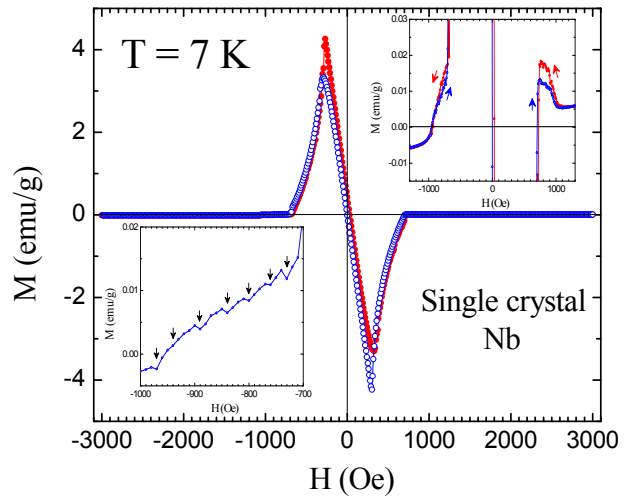}
\caption{\label{fig4}(colour online) Portions of the magnetization vs magnetic field at selected temperatures. In the inset panel, the same curves are shown on an expanded scale and the temperatures are as labeled for different colored curves. The $H_{c2}$ is marked at the crossover field on the linear curve, and $H_{c3}$ is marked at the onet of the divergent behavior.}
\end{figure}

In figure 5, we show the portions of the magnetization hysteresis loops recorded at several temperatures. The sample is cooled in each run from 12 K down to the selected temperature. Cooling field was near zero, and later the the magnetiztion vs field was measured. In the inset we show the magnetified portion close the superconductor to normal transition. In each case, we observe positive magnetization peaks. The $H_{c2}$ is marked at the crossing position of the magnetization curves with the linear paramagnetic bounadry. $H_{c3}$ is marked at the merger position, where the M vs H curve joins the linear paramagnetic state. The positive peaks start to appear at $T$ = 9.0 K, and increases at first with lowering the tempratures. The maximum of the peak height is observed at 8.25 K, and after that the peaks become broader and starts to fall off. This measurement allows us to obtain the values of the $H_{c2}$ and $H_{c3}$ at each temperature below $T_c$. We plot in figure 6, the ($H_{c2}$,$T_{c2}$) and ($H_{c3}$,$T_c3$) lines. Both the lines appear to meet very close to the $T_c$ at near zero magnetic fields. 

Our studies of the dc magnetization in isofield and isothermal modes show two regimes for the observation of the positive magnetization signal in the single crystal Nb sphere. Starting from the negative fields, the change of sign for applied magnetic field does not reverse the sign of magnetization signal at 0.5 Oe. Instead a positive magnetization is observed for the temperature range from $T_c$ down to 2 K. The peak in the positive magnetization at very low magnetic fields $H\leq 1$ Oe also shows the non uniform changes in the magnetization as a function of temperature. It could related to the competition between the positive PME signal and the opposing contribution from the diamagnetic signal when the applied magnetic field is notionally positive. Increasing the magnetic field to 2 Oe and till 4 Oe, the PME signal suppresses within 0.5 K of $T_c$ and the diamagnetic response reaches the Meissner state. At intermediate magnetic fields, from $H \geq 10 Oe$ to 120 Oe, the positive magnetization signal below $T_c$ remains suppressed, and only the diamagnetic signal was seen in the isomagnetic measurements. At higher fields $H > 120 Oe$, the paramagnetic signal resurfaces, and it grows with the increase in the magnetic field. 

\begin{figure}[ht]
\includegraphics[width=0.45\textwidth]{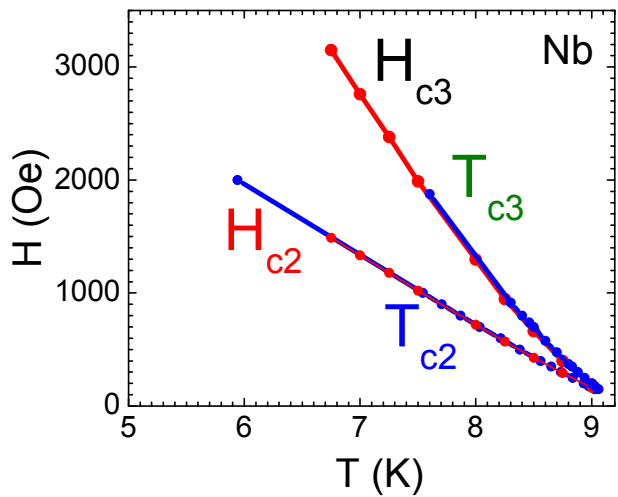}
\caption{\label{fig5}(colour online) $H_{c3}$-$T_{c3}$ phase bouandary in single crystal Nb sphere, along with $H_{c2}$-$T_{c2}$ line.}
\end{figure}

Previous studies in Nb single crystal were attributed the PMFC response to the compression of the magnetic flux trapped within the body of the superconductor. This happens due to the surface layer which is flux free \cite{Saint,Fink} and presence of it can be seen in the ac suscceptibility measurements as well. In single crystal Ca$_3$Rh$_4$Sn$_{13}$ Below 20\,Oe, the surfacing of a curious oscillatory structure in the PMFC response was seen and it was considered as the possible notion of a conservation of angular momentum for the giant vortex state \cite{Moshchalkov, Zharkov} to account for this behaviour. The iso-field and iso-thermal ac susceptibility ($\chi^{\prime}$ and $\chi^{\prime\prime}$) data was shown to register the occurrence of a crossover between the compressed flux regime and the pinned vortex lattice. 

In our isothermal magnetization measurements, the normal to superconductor transition displays the unusual positive response at both sides of the loop. At the notionally negative magnetic fields, the positive response close to the superconductor transition (note the inset in Fig. 4), adds to the diamagnetic response. However, at the notionally positive magnetic field, the superconductor transition is preceded by the positive response, in opposing to the diamagnetic contribution. This response is also seen to be oscillatory and enhances at lower temperatures, suggesting that the further compression of the flux repelled in the interior of the sample due to the nucleation of surface superconducting layer.

To conclude, the magnetic phase diagram shows the plot of $H_{c3}$ line which is slightly concave. The PME signal due to the nucleation of the superconductivity at the surface at very low fields can produce such shape. However, at lower temperatures, the $H_{c3}$ line considerably deviates from the deGennes ratio of $H_{c3}$/$H_{c2}$. Such deviation could be due to additional mechanisms of PME in conventional superconductors and needs further careful measurements in conventional superconductors of different geometries.

\noindent The work form a part of the programme supported by Indo-Spain collaboration ACI-2009-0905. SSB would like to acknowledge the funding from Indo-Spain Joint Programme of co-operation in S \& T, DST, India. We acknowledge the SQUID-VSM facility at TIFR provided by A. K. Grover.

\end{document}